# Comparative Electron Irradiations of Amorphous and Crystalline Astrophysical Ice Analogues

Duncan V. Mifsud[1,2,†], Perry A. Hailey[1], Péter Herczku[2], Béla Sulik[2], Zoltán Juhász[2], Sándor T. S. Kovács[2], Zuzana Kaňuchová[3], Sergio Ioppolo[4], Robert W. McCullough[5], Béla Paripás[6], and Nigel J. Mason[1]

1   *Centre for Astrophysics and Planetary Science, School of Physical Sciences, University of Kent, Canterbury CT2 7NH, United Kingdom*

2   *Atomic and Molecular Physics Laboratory, Institute for Nuclear Research (Atomki), Debrecen H-4026, Hungary*

3   *Astronomical Institute, Slovak Academy of Sciences, Tatranska Lomnicá, SK-059 60, Slovakia*

4   *School of Electronic Engineering and Computer Science, Queen Mary University of London, London E1 4NS, United Kingdom*

5   *Department of Physics and Astronomy, School of Mathematics and Physics, Queen's University Belfast, Belfast BT7 1NN, United Kingdom*

6   *Department of Physics, Faculty of Mechanical Engineering and Informatics, University of Miskolc, Miskolc H-3515, Hungary*

†   Corresponding author:     duncanvmifsud@gmail.com

ORCID Identification Numbers

| | |
|---|---|
| Duncan V. Mifsud | 0000-0002-0379-354X |
| Perry A. Hailey | 0000-0002-8121-9674 |
| Péter Herczku | 0000-0002-1046-1375 |
| Béla Sulik | 0000-0001-8088-5766 |
| Zoltán Juhász | 0000-0003-3612-0437 |
| Sándor T. S. Kovács | 0000-0001-5332-3901 |
| Zuzana Kaňuchová | 0000-0001-8845-6202 |
| Sergio Ioppolo | 0000-0002-2271-1781 |
| Robert W. McCullough | 0000-0002-4361-8201 |
| Béla Paripás | 0000-0003-1453-1606 |
| Nigel J. Mason | 0000-0002-4468-8324 |

**ABSTRACT**

Laboratory studies of the radiation chemistry occurring in astrophysical ices have demonstrated the dependence of this chemistry on a number of experimental parameters. One experimental parameter which has received significantly less attention is that of the phase of the solid ice under investigation. In this present study, we have performed systematic 2 keV electron irradiations of the amorphous and crystalline phases of pure $CH_3OH$ and $N_2O$ astrophysical ice analogues. Radiation-induced decay of these ices and the concomitant formation of products were monitored *in situ* using FT-IR spectroscopy. A direct comparison between the irradiated amorphous and crystalline $CH_3OH$ ices revealed a more rapid decay of the former compared to the latter. Interestingly, a significantly lesser difference was observed when comparing the decay rates of the amorphous and crystalline $N_2O$ ices. These observations have been rationalised in terms of the strength and extent of the intermolecular forces present in each ice. The strong and extensive hydrogen-bonding network that exists in crystalline $CH_3OH$ (but not in the amorphous phase) is suggested to significantly stabilise this phase against radiation-induced decay. Conversely, although alignment of the dipole moment of $N_2O$ is anticipated to be more extensive in the crystalline structure, its weak attractive potential does not significantly stabilise the crystalline phase against radiation-induced decay, hence explaining the smaller difference in decay rates between the amorphous and crystalline phases of $N_2O$ compared to those of $CH_3OH$. Our results are relevant to the astrochemistry of interstellar ices and icy Solar System objects, which may experience phase changes due to thermally-induced crystallisation or space radiation-induced amorphisation.

# 1   Introduction

Laboratory irradiations of low-temperature astrophysical ice analogues using charged particles (i.e., ions and electrons) and ultraviolet photons have demonstrated the formation of complex organic molecules relevant to biology, such as amino acids and nucleobases [1-4]. Such results have highlighted the importance of fully characterising the radiation chemistry occurring within the ice mantles adsorbed on interstellar dust grains, as well as that occurring on the surfaces of outer Solar System bodies such as comets, moons, and Kuiper Belt Objects. To this end, several studies have demonstrated the dependence of this radiation (astro)chemistry on a number of different experimental parameters. Of these, the best studied is perhaps ice temperature, with previous studies documenting the variation in the abundance of molecular products yielded after irradiation at different ice temperatures [5,6].

One experimental parameter which has received considerably less attention is that of the phase of the molecular ice being irradiated. Indeed, to the best of the authors' knowledge, only a handful of studies have directly compared the radiation chemistry between different phases of an astrophysical ice analogue [7-9]. These studies have all considered the electron irradiation of amorphous and cubic crystalline $D_2O$, and have shown that the formation of $D_2$, $O_2$, and $D_2O_2$ was significantly more rapid in the amorphous case. This observation was attributed to the increased ability of radiolytically generated radicals to diffuse through the solid structure in amorphous ices due to their increased structural defects and porosities.

Further understanding the role that the phase of an ice may play in influencing the outcome or productivity of an astrochemical reaction requires that systematic and comparative experiments be performed on a range of molecular ices. Two species which are interesting candidates for such experimental work are $CH_3OH$ and $N_2O$, due to the different intermolecular attractions which characterise their solid phases. $CH_3OH$ is a polar molecule ($\mu$ = 1.69 D) which is expected to form strong hydrogen-bonds. Conversely, $N_2O$ is only anticipated to exhibit weak dipole interactions and van der Waals forces as its main intermolecular interactions in the solid phase, due to its very weakly polar ($\mu$ = 0.17 D) nature. The phase adopted by each molecular ice may have a determining influence on the extent of these intermolecular forces of attraction in the solid ice, which may manifest as differences in the observed radiation chemistry.

$CH_3OH$ and $N_2O$ are also good candidates for such a study since both species are relevant to astrochemistry. $CH_3OH$ ices are ubiquitous in various astrophysical settings, including on Solar System objects such as comets and centaurs [10-12] and around young stellar objects in interstellar space [13-15]. Their presence in such settings is thought to arise from chemical processes within icy grain mantles embedded in dense interstellar molecular clouds, such as the sequential hydrogenation of CO or $H_2CO$ [16-19] or the radiolysis of interstellar or Solar System carbon-bearing water-rich ices [20,21]. $N_2O$ has also been detected in interstellar settings such as stellar nurseries and towards low-mass protostars [22,23]. The presence of $N_2O$ ice, although yet to be confirmed, is expected on the surfaces of outer Solar System moons and dwarf planets, where $N_2$ and $O_2$ ices are known to co-exist in radiation environments driven by giant planetary magnetospheres or the solar wind [24-26].

Although the literature contains a great number of publications concerned with the radiation chemistry of pure $CH_3OH$ and $N_2O$ astrophysical ice analogues [27-38], few make explicit reference to the phase of the ice their experiments are concerned with. This is perhaps unexpected, as prior work has well-documented the phase chemistry of these ices [39-48]. As a solid ice, $CH_3OH$ may exist as an amorphous structure up to a temperature of 128 K [39], after which it undergoes crystallisation. Two crystalline phases are known: a low-temperature α-phase which persists to about 160 K [43,44] and a high-temperature β-phase which persists to liquification at the triple point temperature (175.61 K) [49]. X-ray and neutron diffraction studies have shown that both the α- and β-crystalline phases exhibit orthorhombic symmetry with four molecules per unit cell [41,42,50]. The main distinctions between the phases are the orientation of the hydrogen-bond chains linking adjacent molecules, and the space group to which they belong: the α-phase belongs to the $D_4^2$-$P2_12_12_1$ space group while the β-phase belongs to the $D_{2h}^{17}$-$Cmcn$ space group [43].

Solid $N_2O$ may exist as one of two phases under high-vacuum conditions: an amorphous phase which dominates at temperatures below 30 K and a crystalline phase which exists at higher temperatures and adopts a $Pa3$ structure [47,48]. Interestingly, the phase adopted by a pure $N_2O$ ice is not only determined by the temperature at which it was prepared, but also by its rate of condensation from the gas phase. Hudson *et al.* [47] demonstrated that faster deposition rates are conducive to the formation of a partially or wholly crystalline $N_2O$ ice. The crystalline phase has been described as a 'disordered crystal' due to the fact that the lack of inversion symmetry in the $N_2O$ molecule allows for more than one possible orientation of the weak molecular dipole parallel to the body diagonals of the unit cell. Such a description should not be construed to imply a lack of structure, however, as the crystalline phase is still characterised by an ordering of the constituent $N_2O$ molecules.

To further investigate the possible influence of the phase of a solid ice in determining the outcome of its radiation chemistry, we have performed systematic and comparative irradiations of the amorphous and crystalline phases of pure $CH_3OH$ and $N_2O$ ices using 2 keV electrons. We note that we have not included the β-crystalline phase of $CH_3OH$ in this study due to the high sublimation rates of solid $CH_3OH$ at the temperatures required to synthesise this phase under high-vacuum conditions [44]. We believe that our results constitute the most in-depth analysis of this topic to date.

## 2  Experimental Methodology

Experimental work was performed using the Ice Chamber for Astrophysics-Astrochemistry (ICA) located at the Institute for Nuclear Research (Atomki) in Debrecen, Hungary. The set-up and protocols used have been described in detail in previous publications [51,52], and so only a brief overview is provided here. The ICA is a high-vacuum chamber in the centre of which is a vertically mounted copper sample holder containing a series of ZnSe substrate discs. The sample holder may be cooled to 20 K by a closed-cycle helium cryostat, although the actual temperature may be regulated in the range 20-300 K using a sophisticated temperature control system. The pressure in the chamber is nominally maintained at a few $10^{-9}$ mbar by the combined use of a dry rough vacuum pump and a turbomolecular pump.

The preparation of astrophysical ice analogues on the ZnSe substrates was achieved *via* background deposition by allowing the relevant gas or vapour into a pre-mixing line before dosing it into the main chamber at a nominal pressure of a few $10^{-6}$ mbar through an all-metal needle valve. This deposition pressure was reduced to a few $10^{-7}$ mbar during preparation of the amorphous $N_2O$ samples, due to the known dependence on deposition rate of the phase adopted by this molecular ice [47]. In the case of $CH_3OH$ (VWR, VLSI grade), the liquid was first de-gassed in a glass vial using the standard liquid nitrogen freeze-thaw technique prior to directing the vapour into the pre-mixing line. For $N_2O$, the gas (Linde, 99.5%) was introduced directly into the pre-mixing line without any additional preparation.

Amorphous $CH_3OH$ and $N_2O$ ices were prepared by deposition at 20 K, while their crystalline phases were prepared at 140 and 60 K, respectively. The deposition of these ices could be followed *in situ* using Fourier-transform mid-infrared (FT-IR) transmission absorption spectroscopy. Quantitative measurements made with FT-IR spectroscopy (spectral range = 4000-650 cm$^{-1}$; spectral resolution = 1 cm$^{-1}$) were used to determine the amount of a particular molecular species present within the ice by using Eq. 1 to calculate its molecular column density, $N$ (molecules cm$^{-2}$):

$$N = \frac{P \ln (10)}{A_\nu}$$

(Eq. 1)

where $P$ is the peak area of an absorption band characteristic to a particular molecule (cm$^{-1}$) and $A_\nu$ is the integrated band strength constant for that band (cm molecule$^{-1}$). A list of the characteristic absorption bands and their associated band strength constants used in this study is given in Table 1. From the column density of the initially deposited ice, it is possible to determine the overall ice thickness, $d$ (μm), through Eq. 2:

$$d = 10^4 \frac{NZ}{N_A \rho}$$

(Eq. 2)

where $Z$ is the molar mass of the deposited ice ($CH_3OH$ = 32 g mol$^{-1}$; $N_2O$ = 44 g mol$^{-1}$), $N_A$ is the Avogadro constant (6.02×10$^{23}$ molecules mol$^{-1}$), and $\rho$ is the mass density of the ice, which we have taken to be 0.636 and 0.838 g cm$^{-3}$ for amorphous and crystalline $CH_3OH$, and 1.263 and 1.591 g cm$^{-3}$ for amorphous and crystalline $N_2O$ [47,53]. The constant value 10$^4$ is included in this equation as a conversion factor so as to express $d$ in μm. More information on the initial column densities and thicknesses of the ices investigated in this study may be found in Table 2.

Once deposited, ices were irradiated using 2 keV electrons with projectile electrons impacting the target ices at angles of 36° to the normal over an area of 0.9 cm$^2$. The beam current and profile homogeneity were determined prior to commencing irradiation using the method described by Mifsud *et al.* [52], and an average flux of 4.5×10$^{13}$ electrons cm$^{-2}$ s$^{-1}$ was used. Since the FT-IR spectroscopic beam covered an area of 1.1 cm$^2$, most (> 80%) of the observed ice surface was directly irradiated. Electron irradiations were carried out at 20 K irrespective of the ice phase (i.e., both amorphous and crystalline ices were irradiated at 20 K) so as to exclude any increased mobility of radiolytically derived radicals due to higher temperatures, with FT-IR spectra collected at several intervals throughout the irradiations. As detailed in Table 2, multiple irradiations were performed for each solid ice phase so as to ensure good repeatability of the experiment.

**Table 1** List of characteristic FT-IR absorption bands and their associated band strength constants for the species considered quantitatively in this study. Data collected from various references [34,51,52,54].

| *Species* | **FT-IR Absorption Band (cm$^{-1}$)** | $A_v$ (10$^{-17}$ cm molecule$^{-1}$) |
|---|---|---|
| $CH_3OH$ (amorphous) | 1027 ($v_8$) | 1.61 |
| $CH_3OH$ (crystalline) | 1027 ($v_8$) | 1.28 |
| CO | 2138 ($v_1$) | 1.10 |
| $CO_2$ | 2343 ($v_3$) | 11.80 |
| $H_2CO$ | 1725 ($v_4$) | 0.96 |
| $CH_4$ | 1300 ($v_4$) | 0.80 |
| $N_2O$ (amorphous) | 1283 ($v_1$) | 1.20 |
| $N_2O$ (crystalline) | 1283 ($v_1$) | 0.98 |
| $NO_2$ | 1614 ($v_3$) | 0.62 |
| $N_2O_4$ | 1261 ($v_{12}$) | 0.51 |
| $O_3$ | 1039 ($v_3$) | 1.40 |

CASINO simulations [55] revealed that impinging 2 keV electrons penetrate to a maximum depth of 0.22 and 0.13 μm within the $CH_3OH$ and $N_2O$ ices, respectively (Fig. 1). Given that the thicknesses of the deposited ices were significantly larger than this value, electrons were effectively implanted into the ice. This means that it is true to state that there exists an 'active' zone towards the surface of the ice in which the physico-chemical effects of electron irradiation are manifested. As such, it is often preferable to correct for the 'inactive' zone which is observed spectroscopically but does not play host to any radiation chemistry or physics so as to better illustrate the physico-chemical changes occurring within the 'active' zone of the ice.

**Table 2** List of initial column densities and ice thicknesses of the CH$_3$OH and N$_2$O ices investigated in this study.

| Ice | Species | Phase | $N$ ($10^{18}$ molecule cm$^{-2}$) | $d$ (µm) |
|---|---|---|---|---|
| 1 | CH$_3$OH | amorphous | 1.29 | 1.08 |
| 2 | CH$_3$OH | amorphous | 1.13 | 0.95 |
| 3 | CH$_3$OH | amorphous | 1.17 | 0.98 |
| *Average* | | | *1.20* | *1.00* |
| 4 | CH$_3$OH | crystalline | 1.50 | 1.20 |
| 5 | CH$_3$OH | crystalline | 1.10 | 0.88 |
| 6 | CH$_3$OH | crystalline | 1.09 | 0.87 |
| *Average* | | | *1.23* | *0.98* |
| 7 | N$_2$O | amorphous | 0.29 | 0.17 |
| 8 | N$_2$O | amorphous | 0.28 | 0.16 |
| *Average* | | | *0.29* | *0.17* |
| 9 | N$_2$O | crystalline | 0.58 | 0.27 |
| 10 | N$_2$O | crystalline | 0.61 | 0.28 |
| *Average* | | | *0.60* | *0.28* |

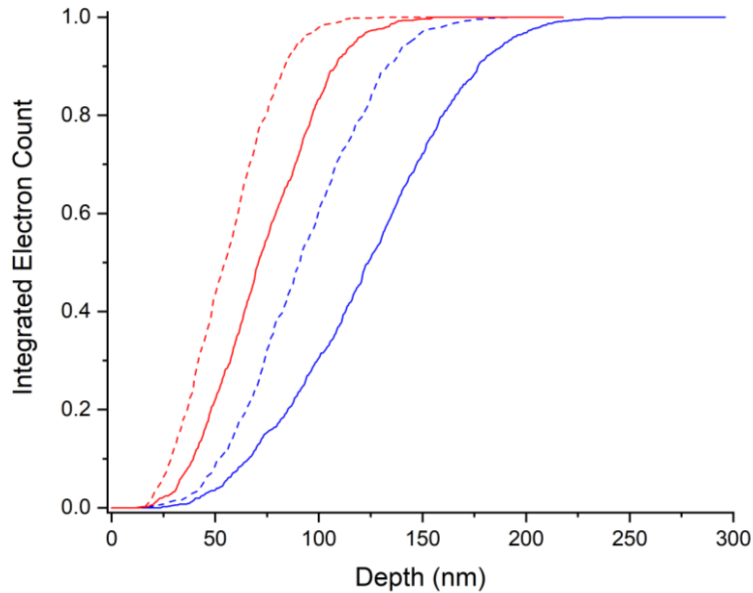

**Fig. 1** Results of CASINO simulations showing the integrated 2 keV electron count as a function of penetration depth into amorphous (solid traces) and crystalline (dashed traces) CH$_3$OH (blue traces) and N$_2$O (red traces) ices. Electrons were simulated to impact the ices at 36° to the normal.

In this study, we have made use of a correction factor when analysing CH$_3$OH and N$_2$O column densities so as to exclude as much as possible the diluting effect of the 'inactive' zones within these ices. To do this, measured column densities were first normalised to the column density measured prior to commencing irradiation. The normalised column density, $N_n$, was then converted to a corrected normalised column density, $N_c$, by means of applying Eq. 3:

$$N_c = \frac{N_n - N_p}{1 - N_p}$$

(Eq. 3)

where $N_p$ refers to the plateau or asymptotic value that the column density of the parent ice reaches after prolonged irradiation as a result of its rates of radiation-induced production and destruction being equal (i.e., the so-called 'steady state'). To calculate $N_p$, we plotted $N_n$ as a function of fluence for each irradiated ice and fitted an exponential decay function to the data. From the equation of this exponential decay function, we calculated the $N_n$ value which would be reached after supplying a theoretical fluence of $1.0\times10^{20}$ electrons cm$^{-2}$ (four orders of magnitude higher than what was actually supplied in this study), and assumed this to be equal to $N_p$. This assumption is justified by the fact that the value for $N_n$ calculated for a fluence of $1.0\times10^{20}$ electrons cm$^{-2}$ (which we have taken to be $N_p$) is equal to that calculated for a fluence of $1.0\times10^{19}$ electrons cm$^{-2}$ for all ices considered, indicating that the steady state would have indeed been reached by this fluence. We wish to draw attention to the fact that this correction factor has only been applied to the $CH_3OH$ and $N_2O$ parent ices, and not to any radiolytically derived product molecules.

## 3 Results and Discussion

### 3.1 Radiation-Induced Decay of Amorphous and Crystalline Ices

In this study, the amorphous and crystalline phases of pure $CH_3OH$ and $N_2O$ astrophysical ice analogues were irradiated using 2 keV electrons at 20 K. As shown in Fig. 2, the radiolytic decay rate of $CH_3OH$ was found to be more rapid in the case of the amorphous ice compared to the crystalline one. Conversely, the radiolytic decay trend of the amorphous and crystalline $N_2O$ ices were observed to be more similar to each other, although that of the amorphous phase was still more rapid than that of the crystalline one. In an effort to quantify the difference in or similarity between the radiolytic decay functions of the irradiated amorphous and crystalline ice phases, the weighted Jaccard coefficient $J_w$ (also referred to as the Tanimoto coefficient) was computed [56,57]. For two real, non-zero measurable functions $f$ and $g$ measured over a given parameter space (in this instance, electron fluence $\varphi$), $J_w$ is given as:

$$J_w = \frac{\int \min(f,g)\,\mathrm{d}\varphi}{\int \max(f,g)\,\mathrm{d}\varphi}$$

(Eq. 4)

As such, $J_w$ varies between 0 and 1, with the latter indicating identical functions and the former indicating no statistical similarity whatsoever. For the interpolated electron-induced decay rate data presented in Fig. 2, $J_w$ was calculated to be 0.40 in the case of $CH_3OH$, and 0.80 in the case of $N_2O$. These results suggest that, although statistical differences do exist between the decay rates of the amorphous and crystalline phases of both molecular species, they are significantly more pronounced in the case of $CH_3OH$ than they are in the case of $N_2O$.

To quantify further, upon supplying an electron fluence of $8.9\times10^{15}$ electrons cm$^{-2}$, the $CH_3OH$ column density in the 'active' zone of the amorphous ice dropped to an average 25% of its original value, while for the crystalline ice this value, on average, only dropped to 67%. This phase-dependent decay of $CH_3OH$ was apparent throughout the irradiation of the ices, with the amorphous ice requiring a fluence of $2.7\times10^{16}$ electrons cm$^{-2}$ for the 'active' zone column density to drop below 10% of its initial value, while triple this fluence was needed to observe this effect in the crystalline phase. Conversely, the average decay trends for amorphous and crystalline $N_2O$ were more similar (Fig. 2), with 44% and 52% of the $N_2O$ in the 'active' zone remaining after a fluence of 1.3×10$^{16}$ electrons cm$^{-2}$ had been respectively delivered to the amorphous and crystalline phases.

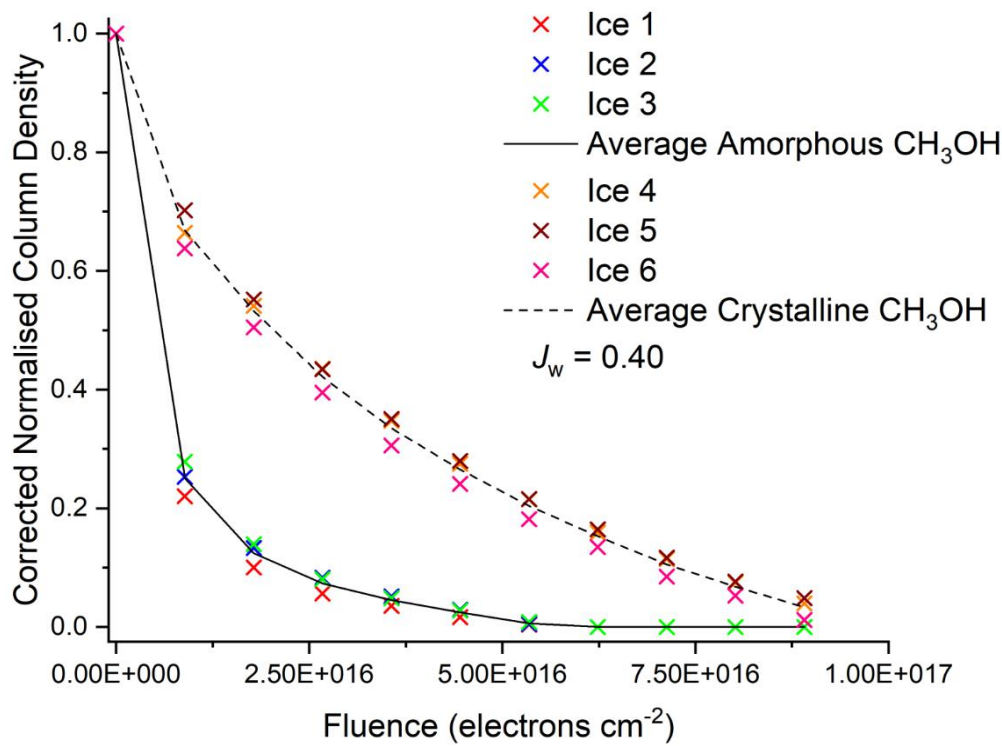

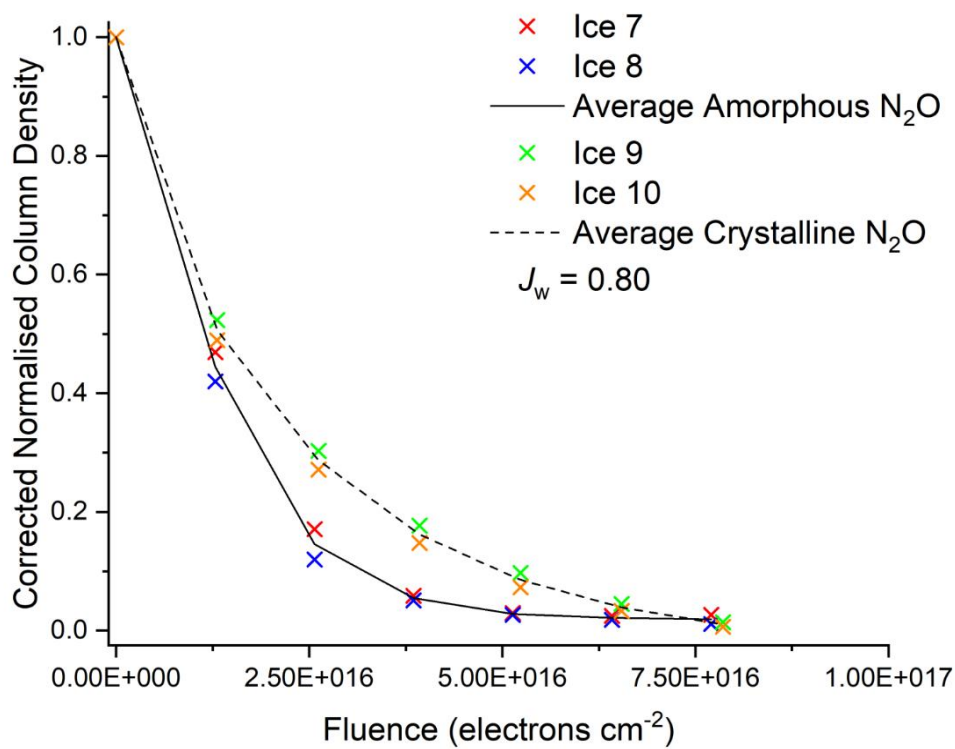

**Fig. 2** Electron-induced decay of CH$_3$OH (above) and N$_2$O (below) ice phases with increasing electron fluence.

These observations may be rationalised in terms of the difference in the nature and extent of the intermolecular bonding present within the different ice phases of $CH_3OH$ and $N_2O$ and, by extension, the stability against radiolytic decay such bonding is able to offer. In the case of $CH_3OH$, the regularity of the crystalline lattice permits an extensive array of hydrogen-bonds which confers an additional degree of stability to the ice. This network of hydrogen-bonds then requires an increased supply of energy (i.e., dose) to be overcome, with some of the incident electrons' kinetic energy being required for this purpose. The amorphous phase is not characterised by this extensive hydrogen-bonding network, and so there is more available energy which may be used for driving electron-induced radiolytic dissociation of this phase, leading to a more rapid initial depletion of $CH_3OH$ parent molecules. It should be noted that, although some hydrogen-bonding between small molecular clusters is expected in the amorphous phase, this would not have the stabilising effect of the extensive lattice found in the crystalline phase, particularly in light of the fact that hydrogen-bonding in $CH_3OH$ is known to be a co-operative phenomenon in which the presence of a hydrogen-bond in the array strengthens subsequent hydrogen-bonds *via* electrostatic effects [58,59].

The need to overcome the hydrogen-bonding network implies that the electron irradiation of the crystalline $CH_3OH$ ice should result in a degree of localised amorphisation of the ice in the 'active' zone into which the projectile electrons penetrate [60]. This is observed in our acquired FT-IR spectra, particularly in the broad absorption band centred at around 3200 $cm^{-1}$ (Fig. 3). In the unirradiated crystalline ice, this band presents as a double peaked structure consisting of a defined, narrow higher wavenumber peak and a broader lower wavenumber peak. On the other hand, this band appears as a very broad single peaked structure in the pure amorphous phase. Upon the onset of electron irradiation of the crystalline $CH_3OH$ ice (and particularly so after fluences of $4.5 \times 10^{16}$ electrons $cm^{-2}$ have been delivered), however, this band is observed to broaden. The effect of this broadening is most visible in the narrower, higher wavenumber constituent peak (Fig. 3) and represents a localised transition to a less ordered ice structure. Once the hydrogen-bonding network has been disrupted, radiolytic dissociation of the $CH_3OH$ molecules may take place more efficiently.

Given such results, the next point to consider is perhaps the possibility of a charged projectile not having sufficient energy to disrupt the hydrogen-bonding network and thus not induce any significant radiolytic dissociation of the $CH_3OH$ crystalline phase that would otherwise occur in the amorphous phase. A similar scenario was considered by Bag *et al.* [61], who demonstrated that $CH_2^+$ molecular ions with kinetic energies of < 8 eV fired towards a $D_2O$ target ice exhibited hydrogen-deuterium exchange reactions when the ice was amorphous, but not when it was crystalline. In terms of electrons as projectiles, although previous studies using lower energies (2.5-20 eV) have been performed, to the best of our knowledge all of these irradiations have been performed on an amorphous target $CH_3OH$ ice [31-33,62,63]. It would therefore be interesting to study the differences between the radiolysis of amorphous and crystalline $CH_3OH$ ices when using a series of lower energy (< 20 eV) electrons with the aim of reproducing the results of Bag *et al.* [61] and of this present study.

The average radiolytic decay trend of amorphous and crystalline $N_2O$ ices were observed to be significantly more similar than were the decay trends of amorphous and crystalline $CH_3OH$, as evidenced by their higher $J_w$ value. Unlike in $CH_3OH$, no hydrogen-bonding network is present in solid $N_2O$. Instead, the dominant intermolecular interactions between adjacent $N_2O$ molecules are weak dipole interactions and van der Waals forces. The large-scale orientation of molecular dipoles in the regular crystalline lattice confers an additional degree of stability against radiolytic decay. However, since dipole interactions are weaker than hydrogen-bonds, this increase in the resistance to radiolytic decay of crystalline $N_2O$ ice compared to its amorphous phase is not as large as that of the crystalline $CH_3OH$ ice compared to its amorphous phase. Once again, we note that some limited clustering due to orientation of dipoles is anticipated in the amorphous phase, particularly due to the known existence of gas-phase $N_2O$ dimers, trimers, and tetramers [64-66]. Such molecular clusters are preserved during condensation and, indeed, the structure of the slipped parallel dimer (or an extension thereof) is thought to be key to the so-called 'spontelectric' (i.e., spontaneously electric) properties of condensed $N_2O$ [67-69].

Our results therefore provide quantitative indications of the increased resistance to radiation-induced decay of crystalline ices compared to their amorphous phases. This resistance is the result of more extensive intermolecular forces of attraction in the regular crystalline lattice which must be overcome before molecular dissociation may take place. Stronger and more long-ranging intermolecular interactions, such as the hydrogen-bonding network in crystalline $CH_3OH$, are able to provide greater resistance than weaker ones, such as the molecular dipole interactions in $N_2O$. As such, we conjecture that the difference in the radiolytic decay trends of amorphous and crystalline ice phases should become negligible for those molecular species which are non-polar and should be more pronounced for those which exhibit very strong and extensive hydrogen-bonding networks in the crystalline phase.

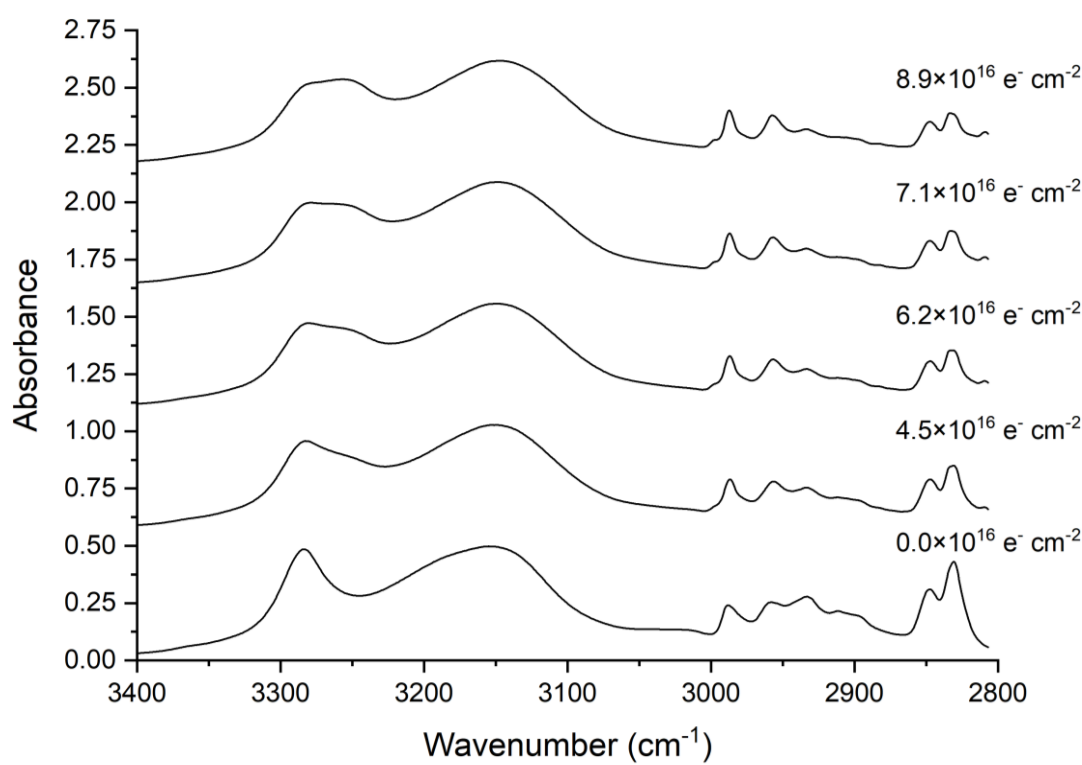

**Fig. 3** The broadening of the double peaked FT-IR absorption band at about 3200 cm$^{-1}$ in crystalline $CH_3OH$ ice as a result of electron irradiation is an indication of localised amorphisation. This broadening is most evident in the higher wavenumber constituent peak. Note that spectra are vertically shifted for clarity.

## 3.2 Radiation-Induced Chemistry

Previous studies have conclusively demonstrated that the chemistry occurring in ices as a result of their irradiation by ions or electrons is in fact driven by a cascade of tens of thousands of low-energy (< 20 eV), non-thermal secondary electrons that are released along the track of the incident charged projectile [62,63,70]. In this study, we have also performed an analysis of the formation of radiation products after the irradiation of amorphous and crystalline pure $CH_3OH$ and $N_2O$ ices. Although we have documented the formation of many new molecular species as a result of the electron irradiation of these ices, we shall limit our discussion to the formation of those species that present well-defined and unambiguous absorption bands in the acquired FT-IR spectra (Fig. 4) and for which the integrated band strength constants are known. In the case of the electron-irradiated $CH_3OH$ ices, these are CO, $CO_2$, $H_2CO$, and $CH_4$, while for the electron-irradiated $N_2O$ ices these are $NO_2$, $N_2O_4$, and $O_3$.

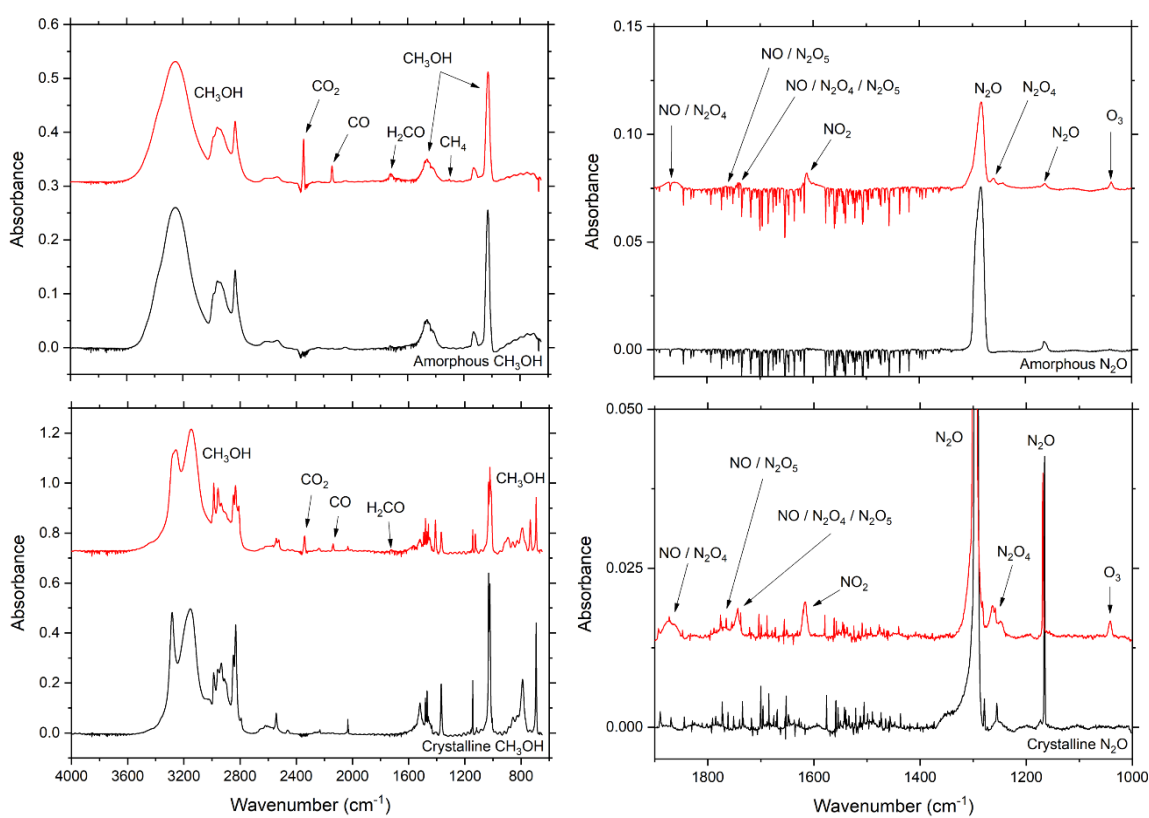

**Fig. 4** FT-IR absorption spectra of the unirradiated (black traces) and 2 keV electron irradiated (red traces) amorphous and crystalline phases of $CH_3OH$ and $N_2O$ ice. In these spectra, the delivered fluences to the irradiated ices were $8.9\times10^{16}$ and $1.3\times10^{16}$ electrons cm$^{-2}$ for the $CH_3OH$ and $N_2O$ ices, respectively. Note also that spectra displayed in the same panel are vertically offset for clarity.

For both molecular species, the formation of products as a result of electron irradiation was greater after irradiation of the amorphous phase compared to the crystalline phase by a factor of between 1.25-3.00 at peak production, depending on the product considered (Figs. 5 and 6). Although it is perhaps intuitive to consider this to be the result of a more rapid dissociation of the parent molecular species in the amorphous phase, an additional contributing cause for this observation are the differences in the porosity and structural defects between amorphous and crystalline ices.

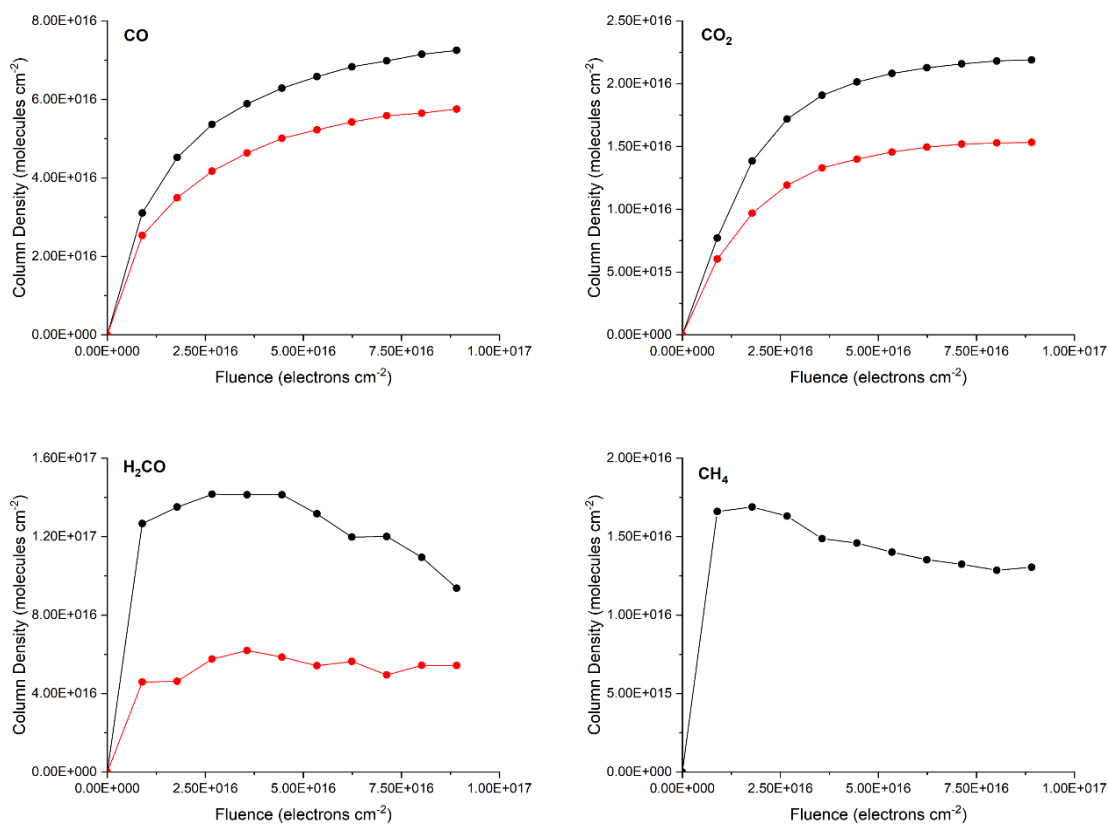

**Fig. 5** Mean spectroscopically measured molecular column densities for the products of amorphous (black traces) and crystalline (red traces) CH$_3$OH irradiation by 2 keV electrons. Although CH$_4$ was detected in our post-irradiative crystalline CH$_3$OH FT-IR spectra, the quantities observed were below the limit of quantitation of our spectroscopic instruments. Note that lines joining the data points are not fits and are plotted only to guide the eye.

At 20 K, radicals formed as a result of the radiolytic dissociation of the parent species are mobile within the ice matrix [71,72], and may therefore diffuse and react with other radicals and molecules to form new species. In amorphous ices, the increase in porosity and structural defects compared to the corresponding crystalline phases allows for these radicals to diffuse more easily throughout the solid matrix [7,73,74]. As such, any recycling reactions leading back to the formation of the parent species are disfavoured due to the necessary radicals having migrated away from each other at the site of their radiolytic formation. With competition from such recycling reactions being lessened, reactions leading to the formation of new product molecules are favoured leading to a greater abundance of these species in the irradiated amorphous phases.

### 3.3 Astrochemical Implications

Our experiments simulate the processing that interstellar icy grain mantles undergo as a result of their interaction with galactic cosmic rays, or of Solar System ices as a result of their interaction with the solar wind or giant planetary magnetospheric plasmas. Such astrophysical environments are known to allow for the cycling of molecular material through different solid phases *via* thermally-induced crystallisation and space radiation-induced amorphisation. For instance, although ices in dense, quiescent interstellar clouds are believed to be principally amorphous, those in stellar accretion discs are more likely to be crystalline [75].

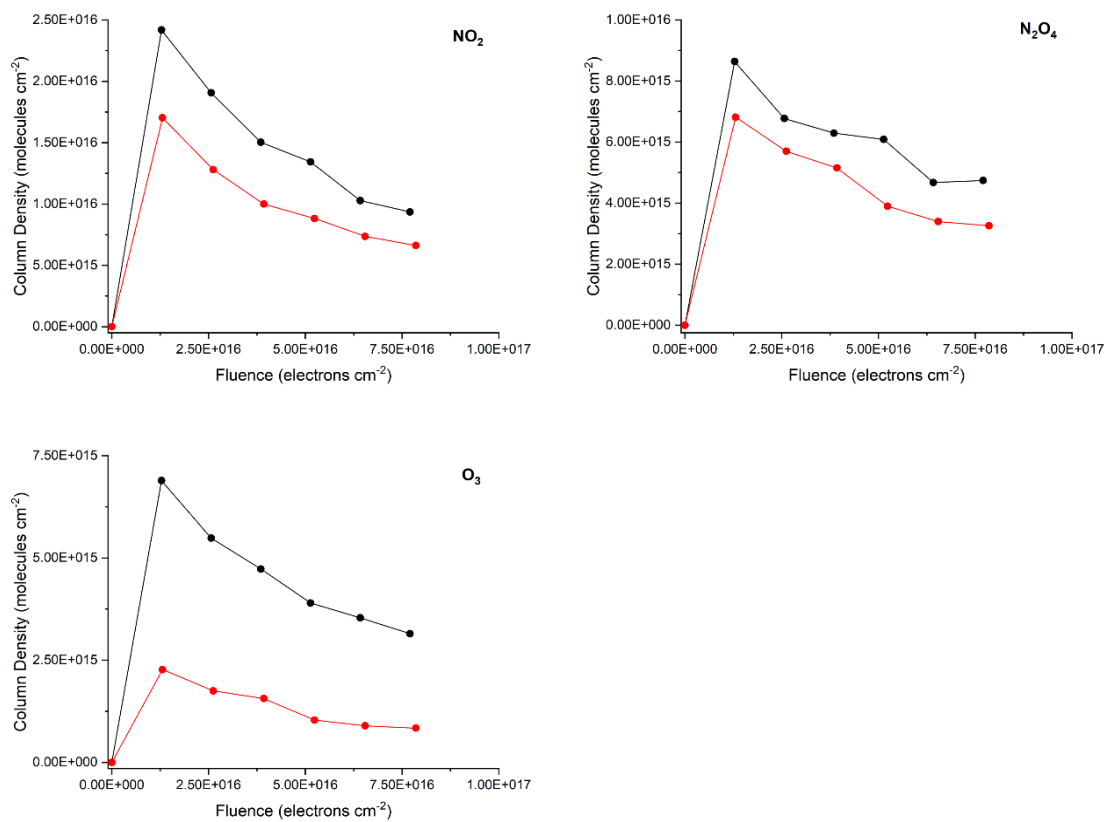

**Fig. 6** Mean spectroscopically measured molecular column densities for the products of amorphous (black traces) and crystalline (red traces) $N_2O$ irradiation by 2 keV electrons. Note that lines joining the data points are not fits and are plotted only to guide the eye.

Our results imply that those ices which are able to form strong and long-ranging intermolecular interactions (such as the hydrogen-bonds in $CH_3OH$) will exhibit a greater resistance to radiation-induced decay in their crystalline phase. Conversely, ices which only exhibit weak intermolecular interactions (such as the dipole interactions in $N_2O$) should exhibit a lesser resistance to radiation-induced decay on transitioning from the amorphous to the ordered, crystalline phase. This interpretation carries implications for assessing the survivability and residence times of different molecular species in various astrophysical radiation environments, and thus highlights the need for both experimental and modelling work in this area of research to take into account the phase of the molecular ice under consideration.

Our results have also demonstrated that, irrespective of the nature, strength, or extent of the intermolecular bonds present within a molecular ice, the amorphous phase is typically more chemically productive than the crystalline one. This has been attributed largely to the increased porosity and structural defects present within the amorphous phase which allow radiolytically generated radicals to diffuse away from the point of their formation more efficiently, thus promoting the formation of new molecules at the expense of the recombination of the parent molecule [7,73,74]. However, our results also suggest that the increased availability of radicals from the faster radiolytic decay rates of the amorphous ices is also likely to contribute to the more rapid accumulation of products in this phase.

The formation of complex organic molecules as a result of the irradiation of an interstellar ice which is rich in species capable of hydrogen-bonding is thus likely to be less productive once the ice has been crystallised as a result of thermal processing (for example, during stellar birth and evolution). As such, the formation of such complex organic molecules may occur to a greater extent in the earliest stages of dense interstellar cloud evolution (i.e., before stellar birth and evolution begin) or in those areas of the dense cloud which are not influenced by the thermal emission of young stellar objects. This idea is not unreasonable, particularly in light of the recent spate of discoveries of complex organic molecules in the pre-stellar cloud TMC-1 [76-79]. It is expected that the recently launched NASA *James Webb Space Telescope* will allow researchers to map out the degree of crystallinity of interstellar and outer Solar System ices in unprecedented detail, thus permitting the influence of solid phase on the outcome of various astrochemical reactions to be further explored [80].

Lastly, we comment briefly on the experimental approach used in this study. This work is a typical example of a one-factor-at-a-time (OFAT) study, in which one experimental parameter (in this case, the phase of the ice under consideration) is varied and the results of such variation on the resultant physico-chemical evolution of the ice is analysed [81]. It has been suggested that such OFAT studies may not be ideal in revealing the true chemical complexity of an astrochemical reaction or process, and that statistically designed experiments and systems science approaches may be more appropriate [82].

Nevertheless, we consider our use of an OFAT approach to be justified, as a great deal of chemical complexity related to the radiolysis of $CH_3OH$ and $N_2O$ ices has already been revealed by previous studies. Moreover, our primary aim in this present study was to determine whether any changes in the physico-chemical evolution of an astrophysical ice undergoing electron irradiation could be induced solely *via* a change in the phase of the ice, thus necessitating the adoption of an OFAT approach. The results of this study confirm the importance of considering the effect of ice phase in determining the outcome of a radiation astrochemistry experiment, and so highlight the need to build this parameter into more exhaustive systems astrochemistry experiments [82].

## 4 Conclusions

We have performed systematic and comparative 2 keV electron irradiations of the amorphous and crystalline phases of pure $CH_3OH$ and $N_2O$ astrophysical ice analogues under high-vacuum conditions at 20 K. Our results have revealed that the rate of electron-induced decay of amorphous $CH_3OH$ is significantly more rapid than that of the crystalline phase, whereas smaller differences could be seen between the electron-irradiated amorphous and crystalline $N_2O$ phases. We have attributed this observation to the fact that the crystalline $CH_3OH$ phase is characterised by an extensive network of strong intermolecular hydrogen-bonds throughout the lattice, which does not exist to the same extent in the amorphous phase. As such, some of the incident electrons' energy must be expended upon first disrupting this network of hydrogen-bonds before driving radiolytic chemistry. In contrast, the weak dipole interactions in $N_2O$ are able to stabilise the crystalline phase to a much lesser degree compared to the amorphous phase, thus leading to a slightly faster radiation-induced decay rate in the latter.

Such results are of direct relevance to astrochemistry, and allow researchers in the experimental and modelling communities to make predictions as to the survivability of various molecules and the rapidity of astrochemical reactions in different astrophysical environments characterised by the competing processes of thermally-induced crystallisation of the ices and their space radiation-induced amorphisation. For example, we hypothesise that other molecular ices capable of forming strong and extensive hydrogen-bonding networks (such as $H_2O$, $HF$, or $NH_3$) should also exhibit slower radiation-induced decay rates in the crystalline phase when compared to the amorphous phase, while non-polar molecules (such as $CH_4$) will show negligible differences in the radiation-induced decay rates of their crystalline and amorphous phases. We believe that the results presented in this study constitute the most in-depth experimental analysis of the role played by the solid phase of an astrophysical ice in defining its radiation physics and chemistry to date, and anticipate that they may act as a good starting point for other studies seeking to further comprehend this topic.

## Author Contributions

The experiment was designed by Duncan V. Mifsud and Perry A. Hailey, and performed by Duncan V. Mifsud, Péter Herczku, Béla Sulik, Zoltán Juhász, Sándor T. S. Kovács, and Béla Paripás. Data analysis was carried out by Duncan V. Mifsud and Zuzana Kaňuchová. All authors contributed to the interpretation of the results, as well as to the preparation of the manuscript.

## Conflicts of Interest

There are no conflicts to declare.

## Acknowledgements

We acknowledge funding from the Europlanet 2024 RI which has been funded by the European Union Horizon 2020 Research Innovation Programme under grant agreement No. 871149. The main components of the experimental apparatus were purchased using funding obtained from the Royal Society through grants UF130409, RGF/EA/180306, and URF/R/191018.

Duncan V. Mifsud is the grateful recipient of a University of Kent Vice-Chancellor's Research Scholarship. The research of Zuzana Kaňuchová is supported by VEGA – the Slovak Grant Agency for Science (grant No. 2/0059/22) and the Slovak Research and Development Agency (contract No. APVV-19-0072). Sergio Ioppolo acknowledges the Royal Society for financial support. The research of Béla Paripás is supported by the European Union and the State of Hungary and is co-financed by the European Regional Development Fund (grant GINOP-2.3.4-15-2016-00004).

We would also like to thank Andrew M. Cassidy (University of Aarhus, Denmark) for the fruitful and stimulating discussions which allowed us to improve our manuscript.